\documentclass[
twocolumn,
elsarticle,
amsmath,
amssymb,
reprint,
floatfix]{revtex4}

\usepackage{srcltx}
\usepackage{natbib}
\usepackage{amsmath}
\usepackage{amsfonts}
\usepackage{amssymb}
\usepackage{amsthm}
\usepackage{newlfont}
\usepackage{tablefootnote}
\usepackage{color}
\usepackage{float}
\usepackage{verbatim}
\newlength{\fw}
\usepackage[pdftex]{graphicx}
\usepackage{subcaption}
\usepackage[pdftex,plainpages=false]{hyperref}
\usepackage{bm}
\usepackage{multirow}
\usepackage{pdflscape}
\usepackage{tabularx}
\usepackage{graphicx}
\usepackage{float}
\usepackage{makecell}
\usepackage[dvipsnames]{xcolor}
\usepackage{physics}
\usepackage{soul}

\newcommand{\twovec}[2]{\left(\begin{array}{c} #1 \\ #2 \end{array} \right)}
\renewcommand{\e}[1]{\scalebox{1.2}{$e$}^{\0#1}}

\begin{document}
\title{Non-uniform Birefringence in Highly-reflective Substrate-Transferred GaAs/Al$_{0.92}$Ga$_{0.08}$As Coatings at 1064~nm}

\author{{Andri M. Gretarsson}\footnote{Corresponding author: \href{mailto:greta9a1@erau.edu}{greta9a1@erau.edu}} }
\affiliation{Department of Physics, Embry-Riddle Aeronautical University, 3700 Willow Creek Road, Prescott, AZ 86301}
        
\author{{Garrett D. Cole}}
\affiliation{College of Optical Science, University of Arizona, 1630 E. University Boulevard, Tucson, AZ 85721}

\author{{Ambroise L.M. Juston}} 
\affiliation{Department of Physics, Embry-Riddle Aeronautical University, 3700 Willow Creek Road, Prescott, AZ 86301}    
\author{{GariLynn Billingsley}, {Camille N. Makarem}}
\affiliation{LIGO Laboratory, California Institute of Technology, 1200 E. California Boulevard, CA 91225}

\author{Benjamin Nicolai} 
\affiliation{Department of Physics, Embry-Riddle Aeronautical University, 3700 Willow Creek Road, Prescott, AZ 86301}
\author{Naomi Borg} 
\affiliation{Department of Physics, Embry-Riddle Aeronautical University, 3700 Willow Creek Road, Prescott, AZ 86301}
\author{{Breck N. Meagher}\footnote{B.N.M. is now at Department of Physics, Syracuse University, 900 S. Crouse Avenue, Syracuse, NY 13244}}
\affiliation{Department of Physics, Embry-Riddle Aeronautical University, 3700 Willow Creek Road, Prescott, AZ 86301}  
\author{{Elizabeth M. Gretarsson}}
\affiliation{Departments of Physics and Aerospace Engineering, Embry-Riddle Aeronautical University, 3700 Willow Creek Road, Prescott, AZ 86301}

\author{{Gregory M. Harry}}
\affiliation{Department of Physics, American University, 4400 Massachussetts Avenue NW., Washington, DC 20016}

\author{{Steven D. Penn}}
\affiliation{Department of Physics, Syracuse University, 900 S. Crouse Avenue, Syracuse, NY 13244 \\ Department of Physics, Hobart \& William Smith Colleges, 300 Pulteney Street, Geneva, NY 14456}

\begin{abstract}
Using a custom-built scanning system, we generated maps of birefringence on reflection at $\lambda=1064$~nm from single-crystal GaAs/Al$_{0.92}$Ga$_{0.08}$As Bragg reflectors (henceforth ``AlGaAs coatings'').  Ten coatings were bonded to fused silica substrates and one remained on the epitaxial growth wafer. The average phase difference on reflection between beams polarized along the fast and slow axes of the coating was found to be $\psi = 1.09 \pm 0.18$~mrad, consistent with values observed in high-finesse optical reference cavities using similar AlGaAs coatings. Scans of substrate-transferred coatings with diameters between 18 and 194 millimeters showed birefringence non-uniformity at a median level of $0.1$~mrad. A similar epitaxial multilayer that was not substrate transferred, but remained on the growth wafer, had by far the least birefringence non-uniformity of all mirrors tested at $0.02$~mrad. On the other hand, the average birefringence of the epi-on-wafer coating and substrate-transferred coatings was indistinguishable. Excluding non-uniformity found at the location of crystal and bonding defects, we conclude that the observed non-uniformity was imparted during the substrate transfer process, likely during bonding. Quantifying the impact on the scatter loss in a LIGO-like interferometer, we find that birefringence non-uniformity at the levels seen here is unlikely to have a significant impact on performance. Nonetheless, future efforts will focus on improved process control to minimize and ultimately eliminate the observed non-uniformity.

\end{abstract}

\maketitle

\tableofcontents
\title{Birefringence Variation in Substrate-transferred AlGaAs Coatings}
\author{Andri M. Gretarsson}
\date{December 2024}

\maketitle

\section{Introduction}

Epitaxial GaAs-based optical interference coatings were first demonstrated by researchers at Bell Labs in the mid 1970s \cite{vanderZiel:75} and are now routinely integrated as mirrors in vertical-cavity surface-emitting lasers, the most widely manufactured class of semiconductor laser diodes \cite{Michalzik2012-ld}. Building on developments in compound-semiconductor micromechanical devices \cite{Hjort1994-pb}, optomechanical resonators fabricated from suspended GaAs/AlGaAs Bragg mirrors were found to exhibit simultaneously low optical and mechanical losses \cite{cole2008monocrystalline}. Owing to their potential for use as low-noise reflectors in ultrastable interferometry~\cite{Harry_text}, a substrate-transfer process was developed using direct bonding to apply AlGaAs coatings to arbitrary optical substrates. Employing a cavity-stabilized laser system locked to an optical reference cavity with AlGaAs coatings, it was shown that the coating elastic loss was reduced by a factor of 10 compared to amorphous oxide coatings~\cite{Cole2013NP}. Over the last decade, continuous refinements have led to the demonstration of ``semiconductor supermirrors" with exceptionally low levels of optical loss in the near and mid infrared, with cavity finesse routinely exceeding 300,000 at wavelengths between 1$\,\mu$m and 5$\,\mu$m~\cite{Cole:16,Truong2023-sx}. AlGaAs coatings are now used in the highest performance ultra-coherent laser systems for optical atomic clocks \cite{JunYe2025} and are expected to improve the astrophysical reach of gravitational-wave detectors~\cite{Fritschell2022,ColeAPL23}. 

First measurements of high-finesse cavities with AlGaAs mirrors showed non-zero static birefringence\cite{Cole2013NP}. This was unexpected given the orientation of the material (100) and the cubic symmetry of the crystal. More recent results in cryogenic cavities have also shown the presence of a noise-like fluctuating birefringence at a similar level as other fundamental noise sources~\cite{Yu2023,Kedar2023}. While this unwanted polarization noise can be minimized by simultaneous locking to and averaging of orthogonal eigenmodes (coherent cancellation of the birefringence fluctuations), it remains a nuisance in cavity-stabilized lasers and may be a technical obstacle to implementation in future gravitational-wave detectors.

There are two primary concerns with the use of AlGaAs coatings in applications such as gravitational-wave detection that require large beam sizes and high optical intensity:
\begin{enumerate}
\item[1]{Noise-inducing interaction between the optical field and the crystal structure and/or other mechanisms for birefringence fluctuation~\cite{Yu2023,Kedar2023,Ma2024,Kryhin2023}}.
\item[2]{Static birefringence in the mirror that is both ubiquitous and shows significant spatial variation.}
\end{enumerate}

Here we report on the latter, with particular emphasis on birefringence non-uniformity, in small (18~mm dia.) through large (194~mm dia.) substrate transferred AlGaAs coatings. We also consider the implications of the observed birefringence non-uniformity for gravitational wave detection.

\section{Experiment Design}

\begin{figure}
\includegraphics[width=\columnwidth]{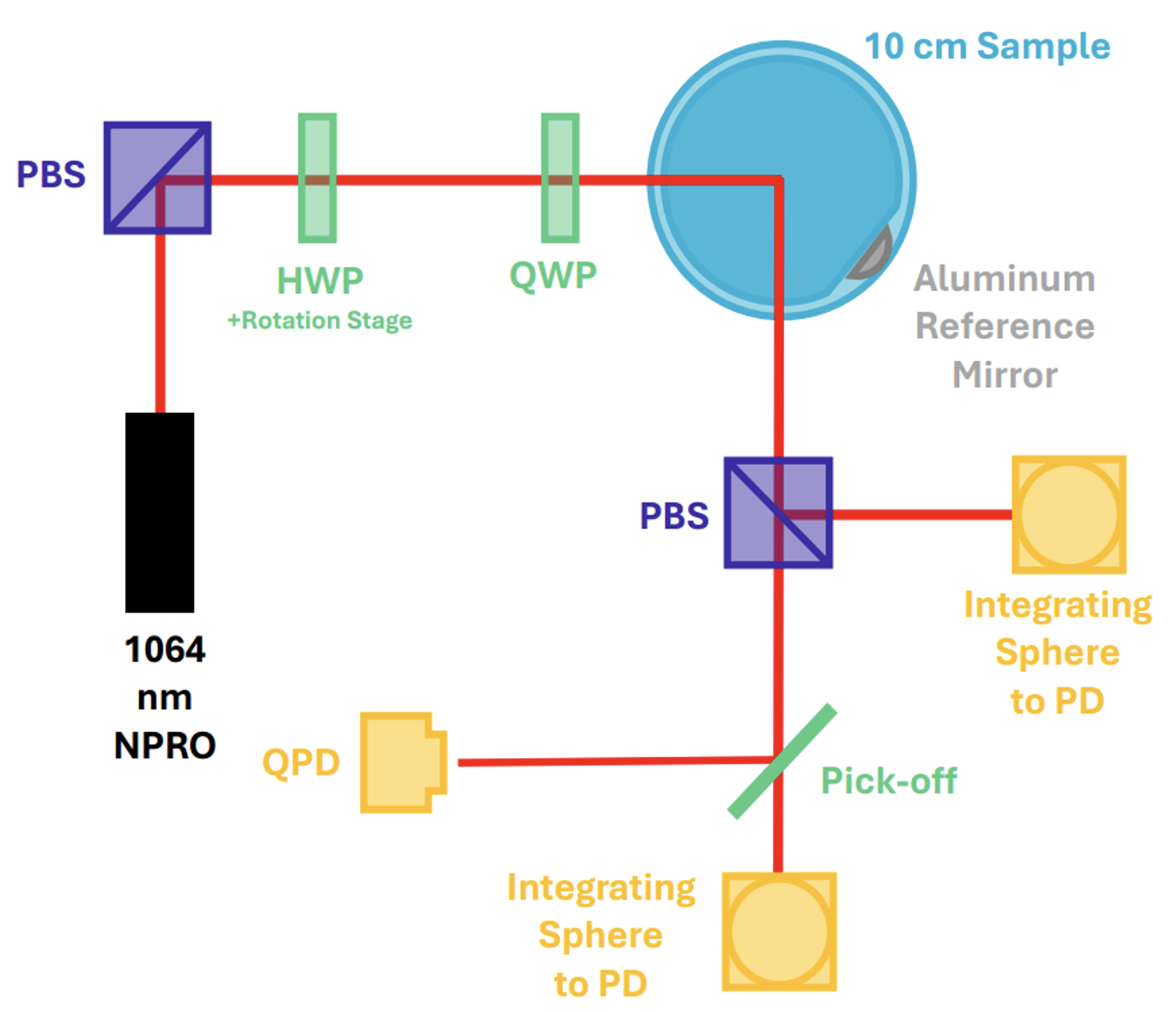}
\caption{Simplified diagram of the setup for measuring birefringence on reflection. (Not shown: Power monitoring pick-off, collimating and focusing optics, steering mirrors, active beam-steering, and rastering $xy$-stage supporting the sample. The beam interrogates the sample at normal incidence. The orientation of the sample is indicated by the ``flat'' (chord) next to the aluminum reference mirror. This was varied, but was typically $\pm45^\circ$, or $\pm135^\circ$ w.r.t. the analyzer s-pol, as shown in the top view here.}
\label{fig:setup}
\end{figure}

We mapped the birefringence in AlGaAs coatings using a single-reflection setup shown in \fig{fig:setup}. Circular polarized light with wavelength $1064$~nm is reflected at normal incidence from the highly-reflective, AlGaAs-coated sample. If the  AlGaAs coating is birefringent, the reflected polarization becomes slightly elliptical. This ellipticity is detected with a polarizing beamsplitter (the ``analyzer'') where the relative power difference in the two polarizations is proportional to the coating birefringence. In order to produce spatial maps of the birefringence, the AlGaAs-coated sample is mounted on a rastering stage that moves the sample under the interrogating beam.

The Jones matrix formalism is convenient for calculating the relationship between the power in the two output polarizations and the mirror birefringence $\psi$. Here, $\psi$ is the difference in the reflected phase due to birefringence for light polarized along the fast axis compared to light polarized along the slow axis. We take the fast axis of the sample to be at an angle $\theta$ relative to $s-$polarization as defined at the splitting surface of the the analyzer. When the beam is traveling parallel to the table surface, $s$-polarization corresponds to the electric field being oriented vertically. The Jones vector corresponding to this input state, immediately before the half wave plate is s-polarized,
\begin{equation}\label{eqn:inputstate}
 	\vec{J_0}\equiv\twovec{J_p}{J_s}=\twovec{0}{1}.
\end{equation}
We multiply this input polarization by the Jones matrices corresponding to the half wave-plate, the quarter waveplate, and the reflective coating which we treat as a waveplate with retardation $\psi$ and orientation $\theta$. This gives the polarization state immediately before the analyzer as
\begin{equation}
	\vec{J}=\twovec{ \frac{1}{\sqrt{2}} ( \e{i\!\frac{\psi}{2}} \cos^2\theta + \e{\uneg\1i\!\frac{\psi}{2}} \sin^2\theta ) -\sqrt{2}\sin\frac{\psi}{2}\cos\theta\sin\theta}
	{\frac{i}{\sqrt{2}}  ( \e{\uneg\1i\!\frac{\psi}{2}} \cos^2\theta + \e{i\!\frac{\psi}{2}} \sin^2\theta ) +i\sqrt{2}\sin\frac{\psi}{2}\cos\theta\sin\theta}.
\end{equation}
The power in the $p$ and $s$ polarizations respectively are given by $P_p=J_p^*J_p$ and $P_s =J_s^{*}J_s$, respectively. With some algebra, we find
\begin{equation}\label{eqn:signal}
\frac{P_s-P_p}{P_s+P_p} = \psi\sin 2\theta.
\end{equation}
The left hand side is the signal we record as a function of position on the coating. Rotating the sample to reveal the sinusoidal dependence of the signal on $\theta$ gives the orientation of the birefringence fast axis. We are primarily interested in maps of non-uniformity, so most measurements are made at $\theta=\pm 45^\circ$ to maximize the signal.  

A significant portion of the development time was devoted to reducing systematic errors in the readout. The main issue was rastering-stage-induced beam motion on the photodiodes and in the beamsplitter. We need to measure relative power changes between the two polarizations on the order of $1\E{-4}$. Since even good-quality photodiodes typically vary by 1 \% or more over their face \cite{Larason1998}, beam motion on the photodiodes induces a strong spurious signal. We added integrating spheres to eliminate this otherwise dominant systematic~\cite{MartyThanks}. To reduce the effect of birefringence in other parts of the optics chain, we actively steer the output beam to keep it in the same position in the optics chain. Coarse and fine actuators steer the final mirror before the sample to correct for beam deviation due to sample curvature or rastering stage tilt. (It should be noted that since actuators are steering the final mirror, the exact position of the beam on the sample is slightly affected by steering. This leads to radial stretching of the image of at most 1\%.) These modifications gave sufficient accuracy to measure the spatially averaged birefringence when a non-birefringent, aluminum-coated reference mirror was used as a control. 

However, we were still limited by instrumental artifacts when mapping non-uniformity. To remove these artifacts, we perform each raster line twice, once with left-circular polarization interrogating the sample, and once with right-circular polarization. This changes sign of the right hand side of \eqn{eqn:signal}, so  subtracting doubles the signal while removing common-mode noise. (The switch in polarization is achieved by rotating the half-wave plate by 45$^\circ$.) Finally, to remove artifacts from presumed gain-drift in the photodiode amplifiers, we fit and removed parabolic features from the individual raster lines. This is roughly equivalent to applying a high-pass filter along the raster lines with a characteristic high-pass length on the order of half the sample size. In other words, our method would not be sensitive to birefringence non-uniformity with lower spatial frequency than about half the sample size that occurs only in the direction of the raster lines. In practice though, this is not a problem because we can check for such non-uniformity by rotating the sample through 90$^\circ$ and comparing maps.

Its worth noting that we also tested the use of a Mach-Zehnder interferometer as a way to read out the phase difference of the two polarizations directly. However, we found that this gave no advantage. Operating mid-fringe, where both output ports pass the same power, the sensitivity to the phase difference $\phi$ between the arms is~\cite{Book:Gretarsson2021}
\begin{equation}
	\frac{P_s-P_p}{P_s+P_p} = \phi
\end{equation}
Since the phase difference between the arms induced by the birefringence is just $\phi=\psi \sin{2\theta}$, the Mach-Zehnder has the same basic sensitivity as the simpler readout we ended up using.

Table~\ref{tbl:samples} lists the samples measured. They consist of small and large samples, and one coating that has not been removed from the epitaxial wafer. All the coatings used were grown in the [100] direction. After transfer to the fused silica substrate, the outwards normal is [$\bar{1}$00]. Some coatings were produced with a straight chord along the edge indicating crystal orientation. This chord corresponds to the intersection of the (1$\bar{1}$0) cleavage plane and the coating surface ($\bar{1}$00).
\begin{table*}
\vspace{3ex}\caption{List of high-reflective AlGaAs mirrors whose birefringence was mapped. Entries with dashes (-) correspond to values that are unknown or not measured for technical reasons. Apart from occasional crystal defects and minor edge damage from the substrate-transfer process, these coatings appear flawless to the naked eye.} \label{tbl:samples}
\begin{tabular}{|l|c|c|c|c|c|l|c|c|}
\hline
     {\bf Sample description} & S/N  &\makecell{\bf{Coating} \\ \bf{dia$\1\1$(mm)}} &\makecell{\bf{Substr.} \\ \bf{dia$\1\1$(mm)}}  &\makecell{\bf{Avg$^{**}$}\\ \bf{ (mrad)}} & \makecell{\bf{Peak-to-valley$^{\dagger}$}\\ \bf{(mrad)}}\\ 
\hline
     Semicircular           &  FS/PL-9102       & 23        & 25.4   &  1.4  & -    \\
     Standard 1"           &  TCS-FS-PL-12264  & 18        & 25.4   &  -    & 0.09   \\
     70~mm, thin             &  3in-TO-Opt/WP    & 70        & 76   &  -    & 0.06  \\
     100~mm, thin            &  132979 004 S     &   89      & 100  &  -    & 0.11   \\
     100~mm, thin, on epi     &  132979 013 S     &   98      & 100  &  0.96 & 0.02  \\
     100~mm, thick (-02)  &  E1800006-02      & 93        & 100  &  1.09 & 0.26  \\
     100~mm, thick (-03)  &  E1800006-03      & 86        & 100  &  1.13 & 0.15  \\
     100~mm, thick (-04)  &  E1800006-04      & 86        & 100  &  -    & 0.10  \\
     200~mm, thin  (A)       &  H8435-068FeF2    & 194       & 200  &  -    & 1.6 \\
     200~mm, thin  (B)       &  1623993-B        & 194       & 200  &  -    & 0.46 \\
     200~mm, thin  (C)       &  P0788-006FEF0    & 194       & 200$^\ddagger$ &  -    & -    \\
     
\hline
\end{tabular}
\vspace{-1 pt}
\begin{enumerate}\setlength{\leftskip}{3 ex}
\itemsep -3 pt 
\item[$\scriptstyle{*}$] \footnotesize{Some coatings did not have a chord indicating crystal orientation. They are indicated by ``-''.}
\item[$\scriptstyle{**}$] \footnotesize{Average birefringence over central reference area compared with a low birefringence control (Protected aluminum first-surface mirror).}
\item[$\dagger$]\footnotesize{The peaks/ridges and valleys of the pattern were traced by eye and marked. The corresponding data was extracted and the mean difference between the high and low regions calculated.}
\item[$\ddagger$]\footnotesize{This sample was broken during manufacture along a straight chord 46 mm from the center.}
\end{enumerate}
\end{table*}

\section{Uniform Birefringence}
\begin{figure}
\vspace{4ex}
\includegraphics[width=1.0\columnwidth]{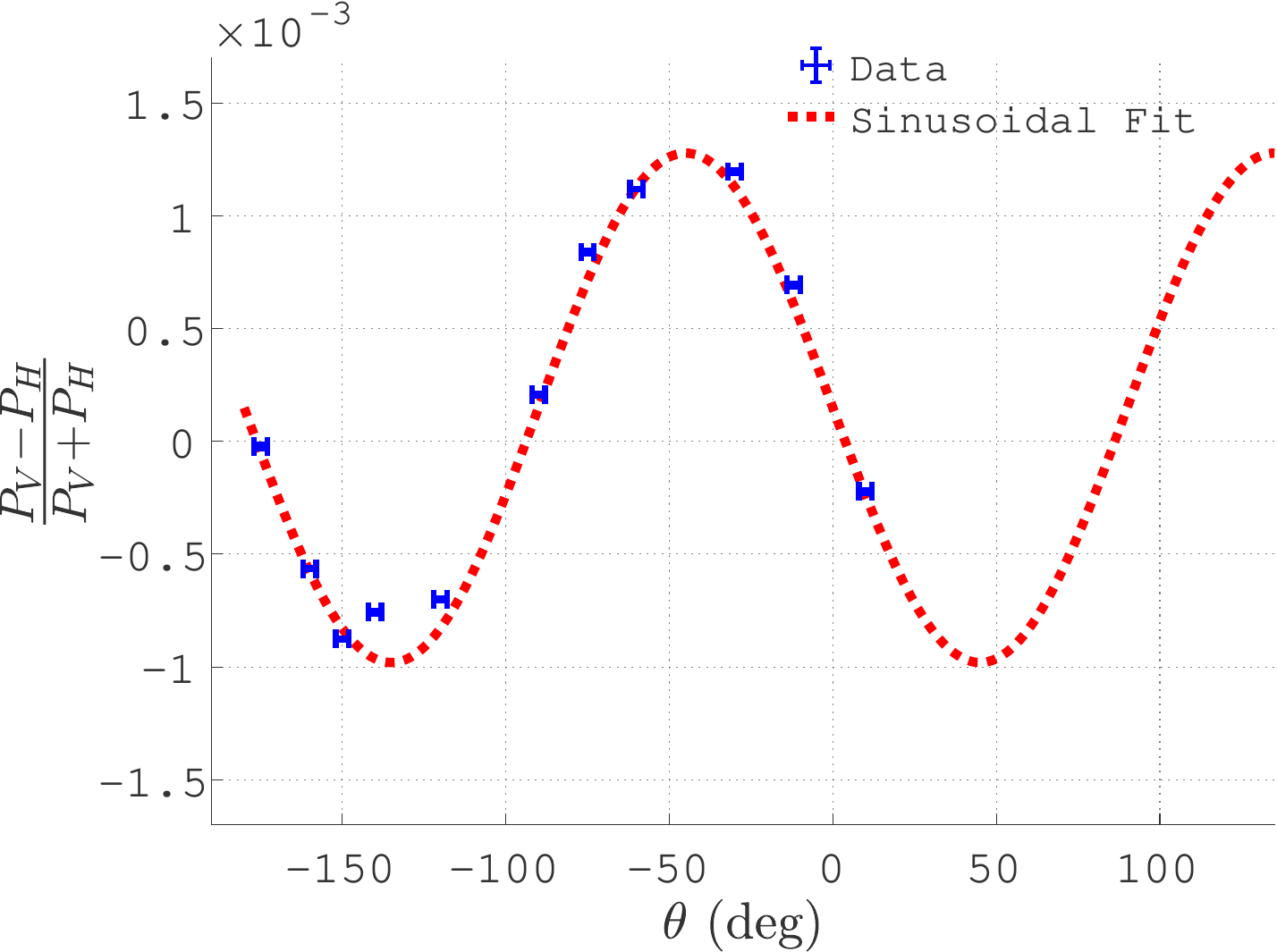}
\caption{Average birefringence of a 93~mm diameter AlGaAs mirror on 10~mm thick silica (S/N: E1800006-02). The magnitude of the birefringence obtained from this data is $1.13\pm0.10$~mrad.}\label{fig:avg_biref_100mm}
\end{figure}

\begin{figure}
\vspace{4ex}
\includegraphics[width=\columnwidth]{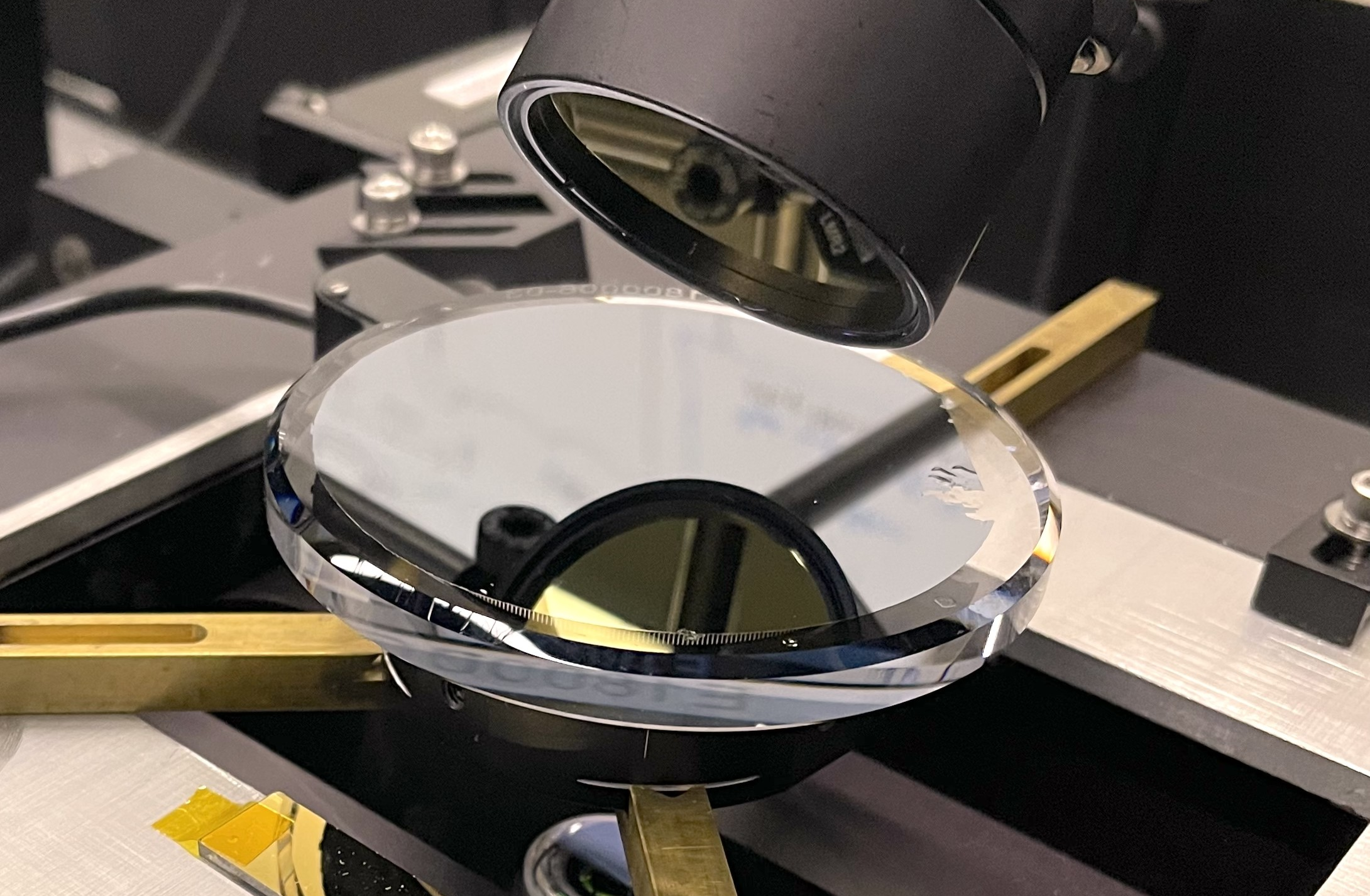}
\caption{A 100~mm$\times$10~mm sample (S/N: E1800006-03) mounted on the rastering stage.}
\label{fig:10cm_sampleA}
\end{figure}

We measured the spatially averaged birefringence in four samples. To do this, we performed multiple scans per sample, rotating the sample about its axis of symmetry between scans. We placed a non-birefringent, Al-coated, protected first-surface mirror next to our sample to act as a control. \Fig{fig:avg_biref_100mm} shows an example of the data for one of our 100~mm$\times$10~mm samples (S/N: E1800006-02) exhibiting the sinusoidal dependence expected from \eqn{eqn:signal}. A similar 100~mm$\times$10~mm sample (S/N: E1800006-03) is shown in \fig{fig:10cm_sampleA}, mounted in the apparatus. The substrate-transferred coatings and the coating that remained attached to the epitaxial growth medium had similar average birefringence. 

\begin{figure}
\vspace{4ex}
\includegraphics[width=1\columnwidth]{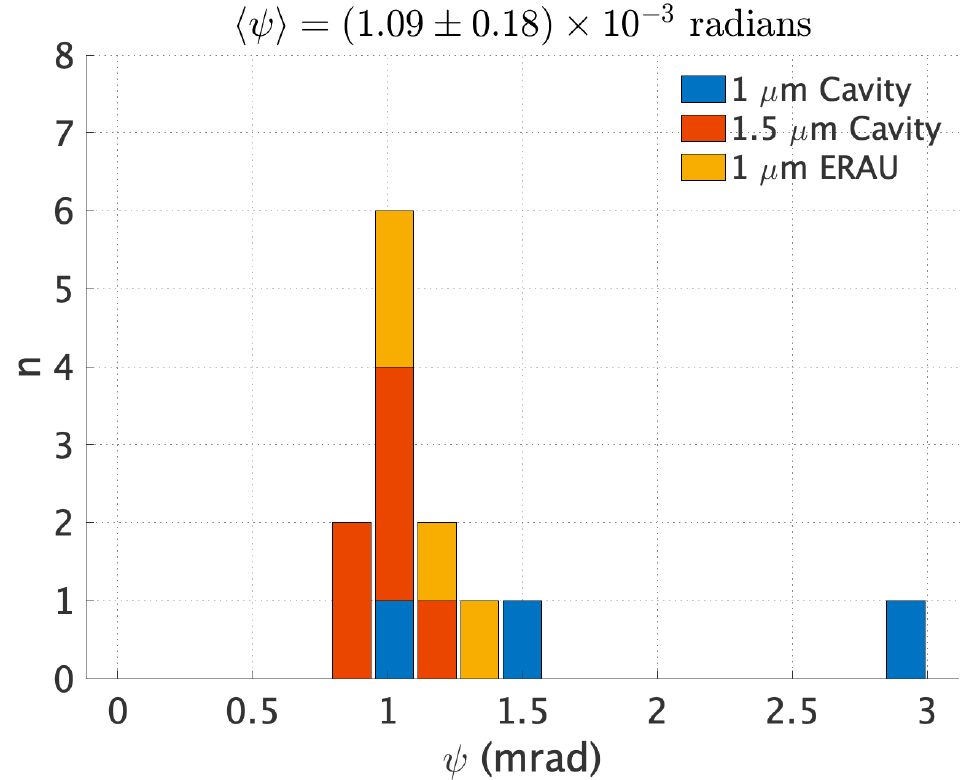}
\caption{Birefringence measured in various cavity experiments compared to the average birefringence measured in our samples. The average birefringence and the RMS deviation from the mean is shown at the top of the figure.}\label{fig:avg_biref_histogram}
\label{fig:avg_biref_hist}
\end{figure}

We compared the average birefringence measured in our samples to the birefringence measured in various cavity experiments at 1.0 and 1.5~${\mu}$m wavelengths. \fig{fig:avg_biref_hist} shows our measurements and the birefringence obtained from mode splitting in several different cavities (collated by one of the authors, GC). Clearly there is significant variation between the birefringence measured across experiments. In cavity experiments, systematic errors may include power-dependent changes in the mode splitting, errors in cavity length, or local sampling of a spatially varying birefringence map. 

The source of non-zero spatially averaged birefringence in AlGaAs coatings may be stress from the 0.2\% lattice mismatch between Al$_{0.92}$Ga$_{0.08}$As and GaAs. However, since the coating material is nominally a pure cubic system, the in-plane strains should be isotropic, precluding strain-induced birefringence. So, if the birefringence is indeed strain-induced, a source of strain anisotropy must be present.  In addition to birefringence, Winkler et al. saw polarization dependence of absorption at 4.54$~\mu$m~\cite{Winkler2021}. They proposed non-uniform strain relaxation in the as-grown epitaxial multilayer as a source of anisotropy (previously seen in InGaP-based materials~\cite{Buckle2018}). Another candidate for strain anisotropy is long-range ordering~\cite{Kuan1985,VANNIFTRIK2006,PRADHAN2024} which could also affect the index directly.

There may be a way to reduce the built-in compressive stress and any associated birefringence by modifying the production method slightly. The full elasticity calculation required to obtain the stress tensor from the lattice mismatch is beyond the scope of this paper, but the following arguments illustrate why the stress is present, that the level should be about $100~MPa$ averaged over the coating, and largely confined to the Al$_{0.92}$Ga$_{0.08}$As layers. To form a continuous crystal, the denser GaAs lattice must stretch, and/or the AlGaAs lattice must compress. During bonding, the thinned remnant GaAs substrate is still much thicker ($\ge100\,\mu$m) than the coating ($\le5\,\mu$m). Therefore, the coating is constrained by the GaAs and it is primarily the Al$_{0.92}$Ga$_{0.08}$As layers that will be compressed to match the GaAs lattice. The remnant GaAs substrate is not etched off until after transfer and bonding to the final substrate, so the coating never relaxes. The constraining tension is just passed from the epitaxial wafer to the final substrate. (When the substrate is thin, the compressive strain in the coating is evident by significant bowing of the substrate.) In any case, we expect compressive strains on the order of $\epsilon_{i}\sim2\times10^{-3}$ in the AlGaAs layers, where $\epsilon_{i}$ are the first three components of the strain in Voigt notation. Using the inverse of the stiffness tensor for orthotropic materials~\cite{Boresi1993}, and ignoring Poisson's ratios, it's straightforward to see that the corresponding stresses will have order of magnitude $\sigma_{i}=E_i\epsilon_{i}$ (no summation implied) where $E_i$ are the Young's moduli in the coordinate directions in the plane of the coating, assumed to be aligned with the cubic crystal structure. The $E_i$ are on the order of 100~GPa~\cite{Adachi1985}, so we expect compressive stresses on the order of 200~MPa in the Al$_{0.92}$Ga$_{0.08}$As layers only. The average stress in the coating is about half that, 100~MPa,  since the GaAs layers constitute roughly half of the coating thickness. 

Based on this, it might be possible reduce the stress in the AlGaAs layers by about a factor of two by releasing the coating from the growth wafer prior to bonding. This would transfer some of the stress to the GaAs layers but with the opposite sign. If the coating were free from the GaAs growth wafer, each layer material would have a strain corresponding to approximately half of the lattice mismatch since the elastic moduli of GaAs and Al$_{0.92}$Ga$_{0.08}$As are similar. The opposing strains in alternating layers could lead to partial cancellation of the average birefringence. Further reduction may be possible through careful coating design.

\section{Non-uniform Birefringence}
\label{sec:SpatialVariation}
The main motivation for the measurements described in this article was to look for the presence of non-uniform birefringence in AlGaAs coatings. Methods for reducing the effect of \textit{uniform} birefringence in optical cavities--90$^\circ$ relative rotation of the cavity mirrors' fast axis, and/or careful alignment of the optical polarization with an axis of the birefringence--may not be effective for non-uniform birefringence. Therefore, we must characterize the amount of non-uniform birefringence and estimate its effects.

\begin{figure}
        \centering
        \includegraphics[width=1\linewidth]{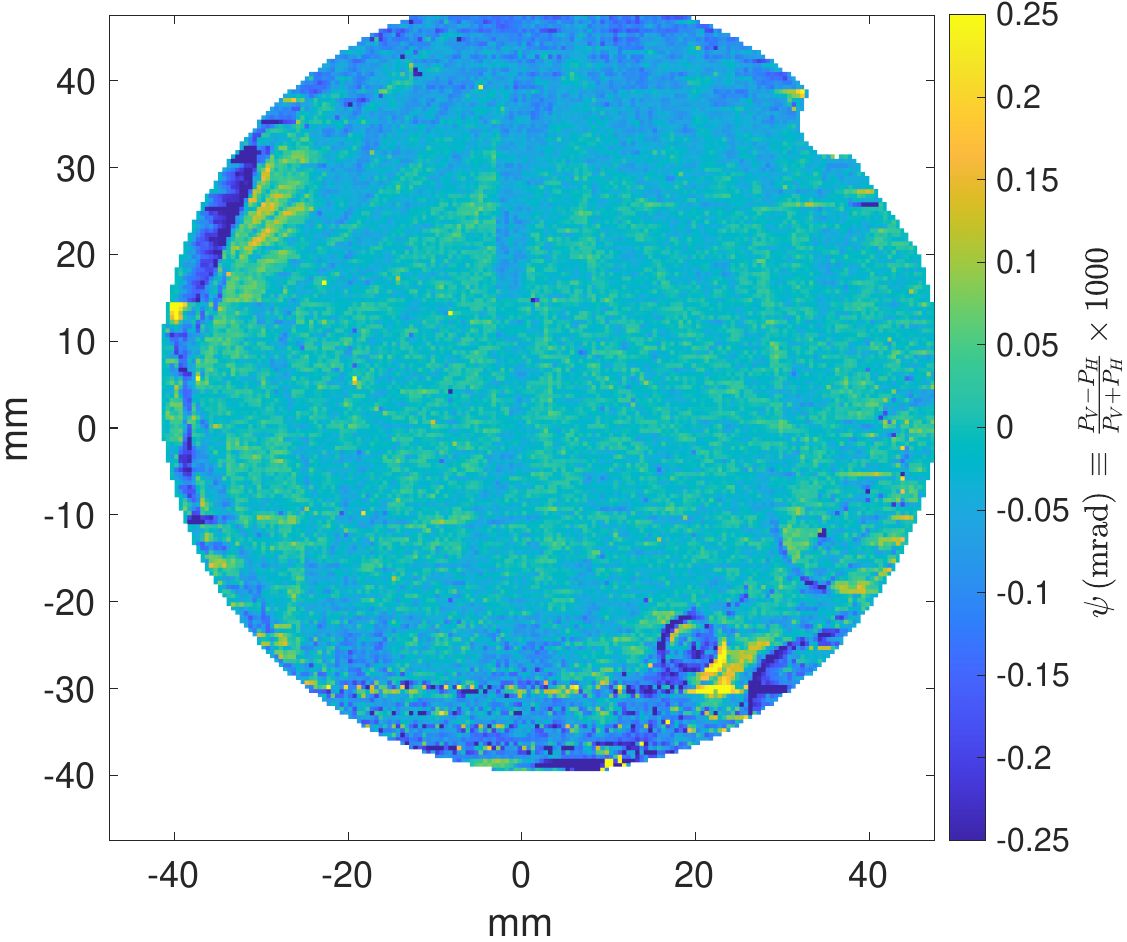}
        \caption{Birefringence variation in a 93~mm AlGaAs coating on 100~mm diameter by 10~mm thick fused silica substrate, S/N: E1800006-02. Fast axis of the average birefringence is at 45$^\circ$ to the analyzer.}
        \label{fig:10cm_variation_aligned}
    \end{figure}

\begin{figure}
        \includegraphics[width=1\linewidth]{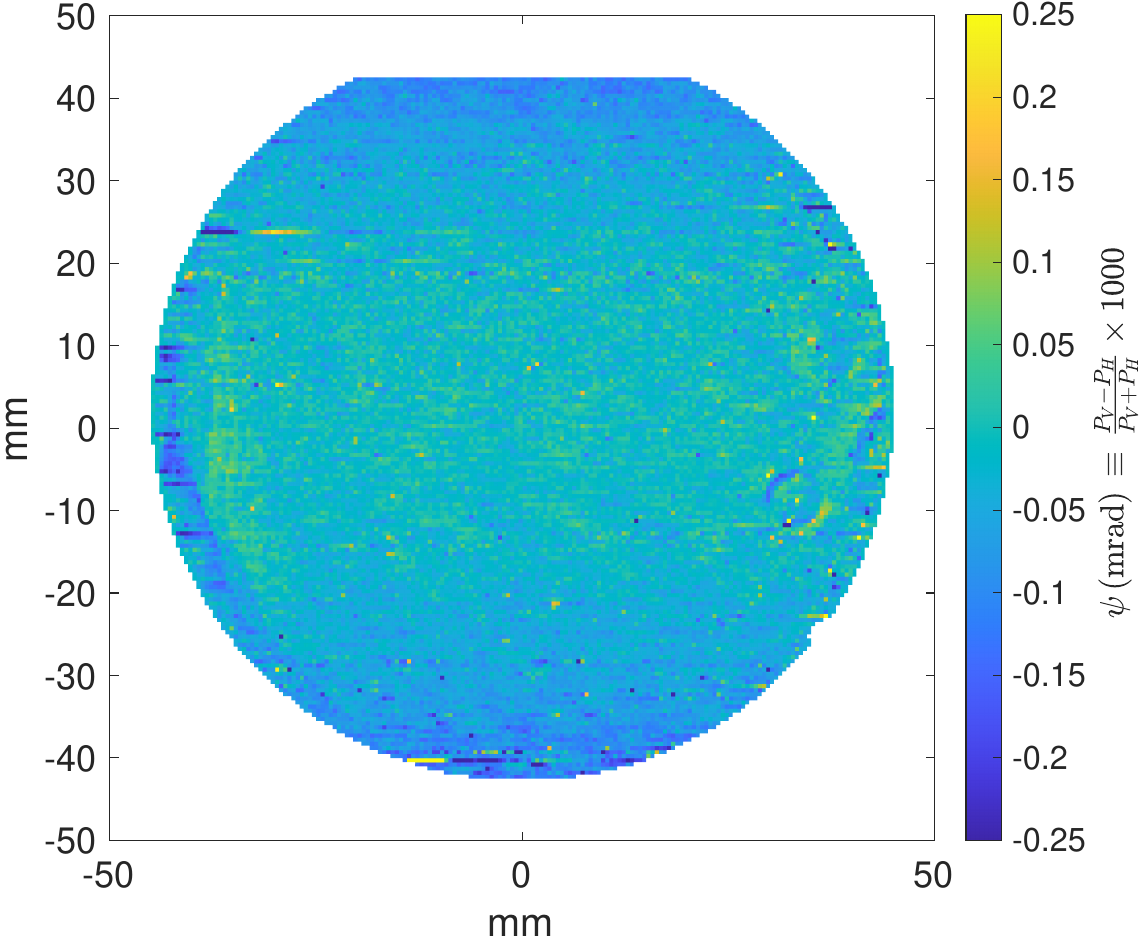}
        \caption{In this measurement, the sample has been rotated so that the fast axis of the average birefringence is parallel to the analyzer (s-polarization). The birefringence variation has largely disappeared, showing that the variation and the average birefringence are similarly oriented.}
        \label{fig:10cm_variation_unaligned}
\end{figure}

We mapped the birefringence variation in 11 coatings of various sizes between 18~mm and 194~mm in diameter. Since coatings for future gravitational-wave detectors will need to be at least 300~mm in diameter, we are particularly interested in large coatings. \fig{fig:10cm_variation_aligned} shows typical birefringence variation in a large coating (S/N: E1800006-02). For this measurement, the fast axis of the average birefringence is at 45$^\circ$ to the analyzer's s-polarization. Clearly, there is a non-random structure to the variation. The level of the peak-to-peak birefringence variation in the center of the sample is on the order of 0.1~mrad, which is about 10 \% of the average birefringence, but larger by a factor of two or more towards the edge. Rotating the sample by 45$^\circ$ so that the fast axis of the birefringence is aligned with the analyzer (\fig{fig:10cm_variation_unaligned}) we see that the structure almost disappears. The non-uniform birefringence is therefore a modulation in the amplitude of the average birefringence, but does not seem to affect the direction of the fast/slow axes much. Exceptions occur near crystal defects like the ``crater-like'' defect seen at the top-right in \fig{fig:10cm_variation_aligned} and mid-right in \fig{fig:10cm_variation_unaligned}. It does not disappear like the more distributed pattern, suggesting that the orientation of the birefringence induced by the defect is different from that of the average birefringence.

To check whether the birefringence was the same on both sides of the coating, we repeated some measurements with the sample placed upside down, with the coating side on the translation stage and the laser entering from the substrate side. Because the silica substrate is transparent to the laser, we can measure the back side of the coating. We found the same pattern of spatial variation on both sides of the coating, which is consistent with a coating-strain-induced birefringence variation.

\begin{figure}
  \centering
  \subfloat[Substrate transferred.\label{fig:a}]{%
    \includegraphics[width=0.463\linewidth]{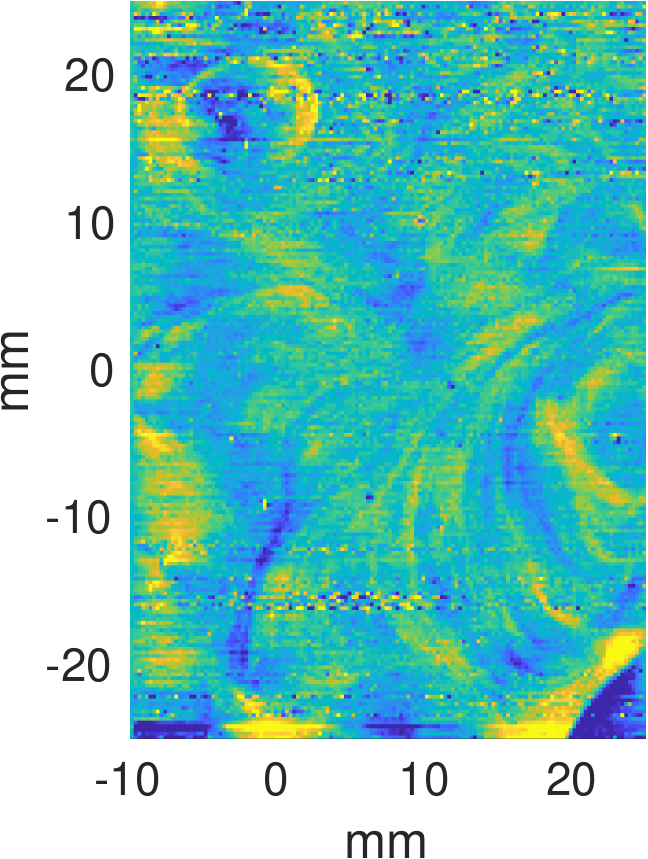}}\hfill
  \subfloat[Still on epi wafer\label{fig:b}]{%
    \includegraphics[width=0.5365\linewidth]{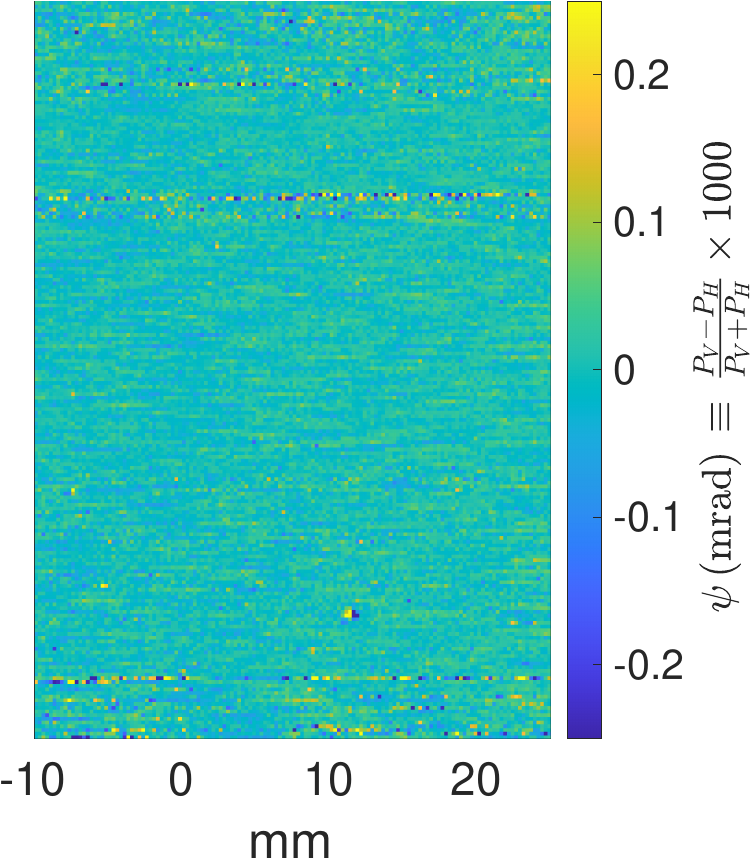}}
    \caption{Central regions of two similar $\sim100$~mm coatings. The coating on the left (S/N 132979 004 S) has been transferred to a silica substrate, whereas the one on the right (S/N 132979 013 S remains attached to the epitaxial growth wafer. }\label{fig:transferred_vs_not}
\end{figure}

To check whether the birefringence was due to the substrate transfer process, we compared the birefringence variation between a substrate-transferred coating and a similar coating that was never removed from the epitaxial growth wafer. (See \fig{fig:transferred_vs_not}).  The coating that remained on the growth wafer showed average birefringence of $\psi=0.96\pm0.1$~mrad, similar to the average birefringence in substrate transferred coatings. However, the coating attached to the growth wafer showed completely uniform birefringence within our sensitivity, except at the locations of a few small defects (possibly dust contamination). We conclude that other than near crystal defects, birefringence non-uniformity seen in AlGaAs coatings is due to the substrate-transfer and bonding process.

\begin{figure*}
    \centering
    \includegraphics[width=1\linewidth]{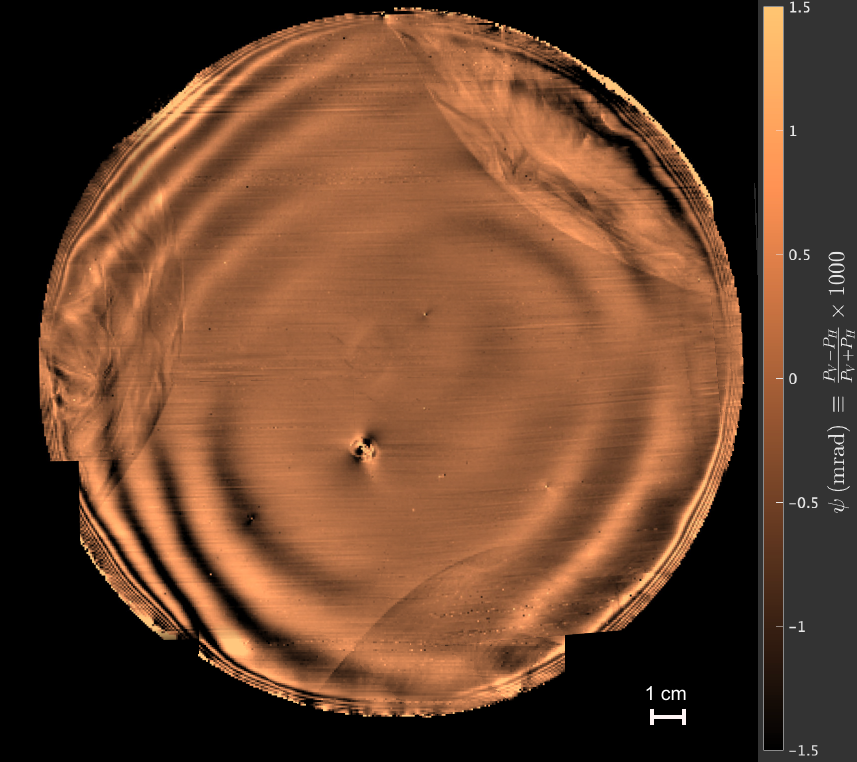}
    \caption{194 mm AlGaAs on 200 mm substrate exhibiting three types of birefringence non-uniformity: A crystal defect just below and left of center, three mandorla-shaped regions of rapidly varying birefringence, and large concentric rings.  This is a composite map made by combining maps of individual sub-regions. (Due to limited xy-stage range, the 200~mm diameter sample could not be mapped in one step.)}
    \label{fig:SampleA}
\end{figure*}

\fig{fig:SampleA} shows the birefringence variation in a 194 mm diameter AlGaAs coating (SN: H8435-068FeF2). The sample is oriented so that the fast axis of the average birefringence is at 45$^\circ$ to the analyzer to maximize the signal; however, in the figure, the average birefringence is filtered out, leaving only the variation. This 200~mm sample exhibits relatively large birefringence variations in places, on the same order of magnitude as the average birefringence. We assume that like the average birefringence, the birefringence variations are strain related. As in other samples, there appear to be two sources of strain variation: variation due to bonding and variation associated with a large crystal defect slightly below and left of center. 

During bonding, the bond is initiated at one location, often the center of the coating, and travels in a ``bond wave'' across the sample until the coating is fully attached. (In everyday life, we may see a similar bond wave when installing a glass cell phone screen protector.) To obtain a strain-free coating, the bond wave must contact the substrate in the same way as a rolling rigid cylinder. Since the the coating bonds on contact, and cannot slip to even out in-plane stresses, stress is incorporated into the coating whenever it doesn't roll on smoothly. For example, the three mandorla-shaped regions around the edge of the sample shown in \fig{fig:SampleA} exhibit rapid strain variation that's ``frozen in.'' The mandorla shape is presumed to be caused by slight drooping of the coating as it hangs between three support tabs (``separation flags'' in the commercial wafer bonding tool). As the bond-wave approaches from the center, the coating---drooping between the supports---is not the correct shape to roll on like a rigid cylinder, leaving a characteristic pattern of rapidly varying strain.  A similar bond-induced non-uniformity is evident in the irregular rings centered on the middle of the coating. There is a faint ``puddle-like'' region just to the left of center where the coating appears to have first made contact with the substrate as it drooped slightly. The bond wave most likely traveled outward from there in an expanding ring. The ring-like ``ripples'' are presumably due to the inability of the coating to relax from the slightly stretched sagging shape to the final, flat configuration.  Interestingly, these rings were absent (or at least greatly reduced in amplitude) in two more 200~mm AlGaAs coatings. In at least one of them, the bond was known to have initiated along an edge. All of these coatings were supported by three tabs and as a consequence showed the mandorla shapes. The last of the three 194~mm coatings (P0788-006FEF0) had the flags pulled out as the bond wave approached the edge which served to reduce the amplitude and spatial frequency of the variations in the mandorla-shaped regions.

The strain pattern induced by bonding can be expected to vary according to the precise support method, location of initial contact, and consequent progression of the bond wave. We see this quite clearly in the difference between the birefringence variation induced in different regions of the 194~mm samples and in the differences between the large and smaller samples. Regardless, birefringence non-uniformity was apparent in \textit{all} the substrate-transferred AlGaAs coatings we mapped.

\begin{figure}[htb]
  \centering
     \subfloat[Background image shows birefringence. Foreground image, inset, shows a height map of the central region. The RMS height variation is $\sim 0.2-0.6$~nm (except for a defect in the center of the coating). The color bar on the right in the indicates the birefringence variation of the background image.\label{fig:a}]{%
     \includegraphics[width=1.0\linewidth]{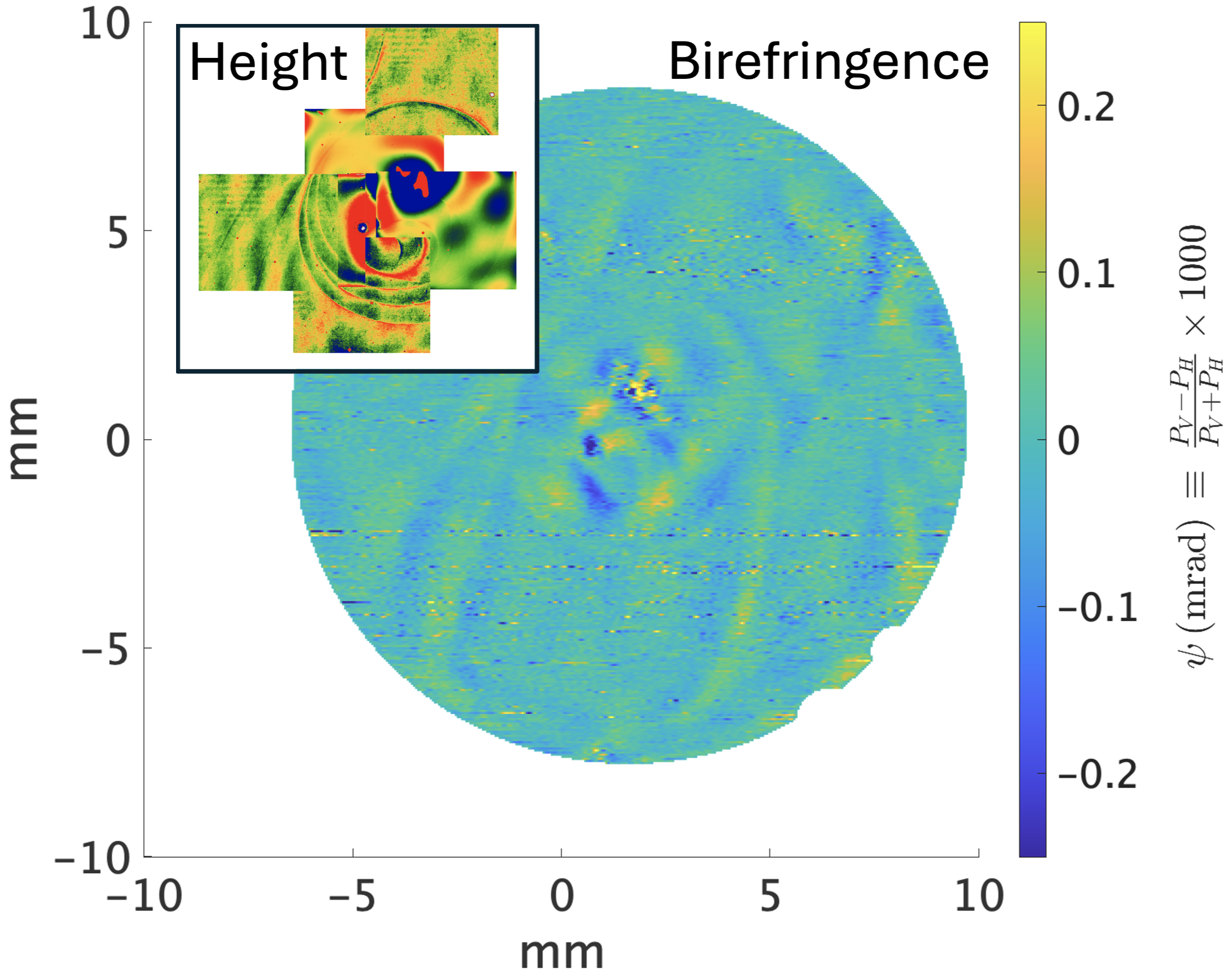}}\hfill
     \subcaptionbox{Overlay of the two data sets.  The geometrical features of the two maps are strikingly similar. When the two images are overlaid, the congruence is evident. To make features in both data sets visible, the height image was desaturated and placed behind the birefringence image which was made slightly transparent. The image is zoomed slightly to show the region of congruence.\label{fig:b}}[\linewidth]{%
     \includegraphics[width=0.65\linewidth]{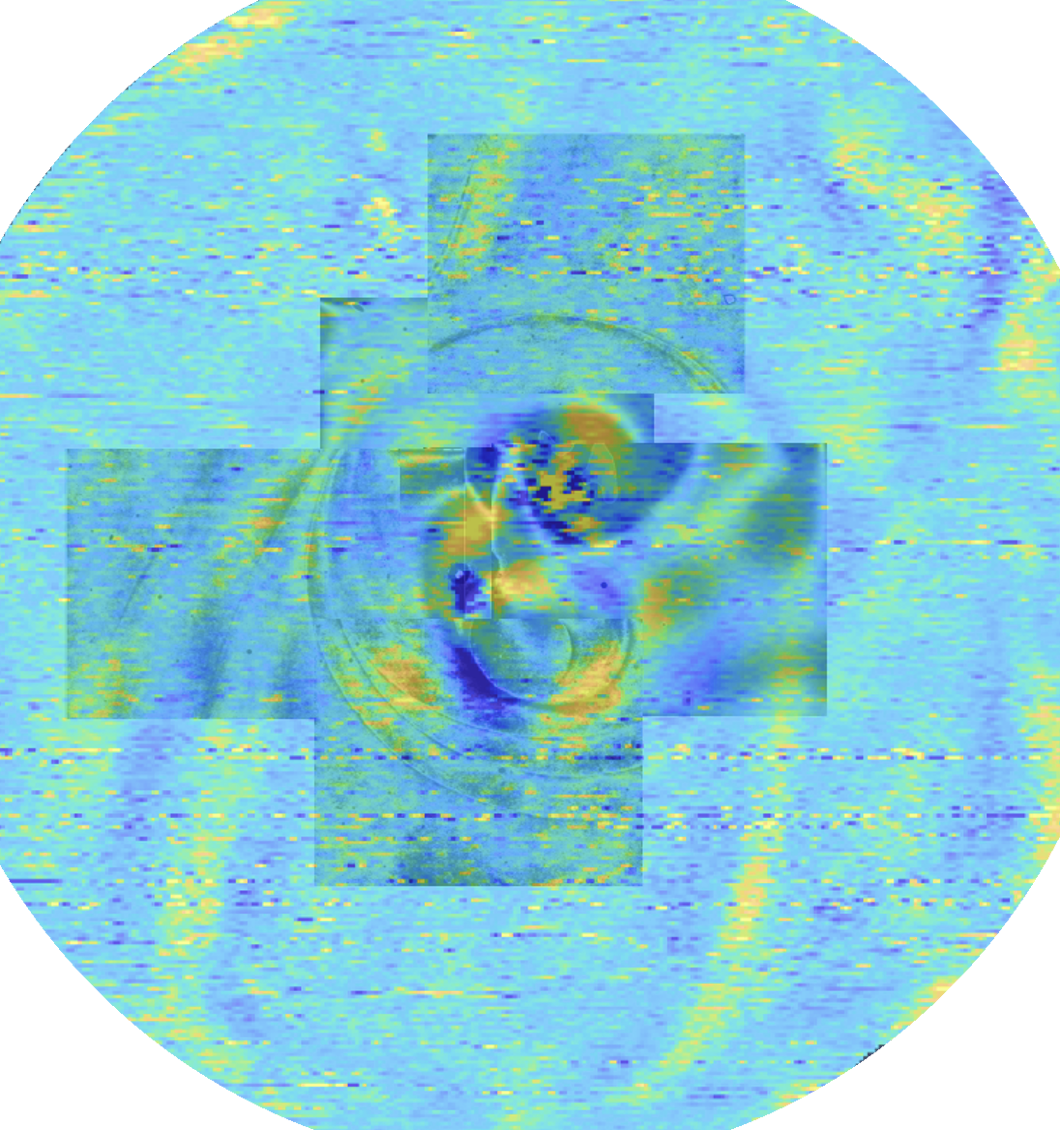}}
    \caption{Comparison between surface height measurements and birefringence for the 18~mm diameter sample, TCS-FS-PL-12264.}\label{fig:roughness_correlation}
\end{figure}

The smallest sample was sent to the LIGO Laboratory at Caltech for surface metrology measurements (performed by one of the authors, CM). The results showed clear correlation between the birefringence variation and sub-optical surface height variations. (See \fig{fig:roughness_correlation}). The RMS surface height variation was approximately 0.2~nm everywhere except for the very central region where there is a small defect. The strain variation is assumed to deform the coating slightly during the bonding process, resulting in increased bulging, ridges, etc. The congruence between the surface height and birefringence patterns supports the assumption that strain variation is the cause of the birefringence non-uniformity. 

\section{Optical Loss in the LIGO $A^{\#}$ Arm Cavities}
\label{OpticalCavities}
AlGaAs coatings are a candidate for a future upgrade to the current LIGO detectors, such as LIGO~$A^\#$~\cite{postO5report}. Therefore, the effects of birefringence on the detector operation must be understood and factored into the design. Uniform birefringence may lead to increased phase noise through any mechanism that drives temporal fluctuations in the birefringence. Examples include mechanical vibrations, coupling with a thermodynamic Langevin force, piezoelectric coupling to fluctuating electric fields, fluctuations due to optical driving of free carriers, etc. These potential noise-couplings are currently being explored by several groups, analytically, through numerical simulations, and experimentally~\cite{WINKLER1994,Kryhin2023,Yu2024,Zhang2024,Wu2024,Gretarsson_DAWN_2024}. 

Here, we use the non-uniform birefringence data presented above to estimate the expected optical loss in a LIGO arm cavity. This assumes no improvement beyond the non-uniformity seen in current AlGaAs coatings. The method only captures the effect of phase variations associated with anisotropic index variation--i.e. associated with birefringence--but not  from  the isotropic index variations. Therefore, this approximation is useful insofar as we assume that isotropic variations of the index are similar to, or smaller than, the anisotropic variations we detect via birefringence.
\begin{figure}
    \centering
    \includegraphics[width=0.95\linewidth]{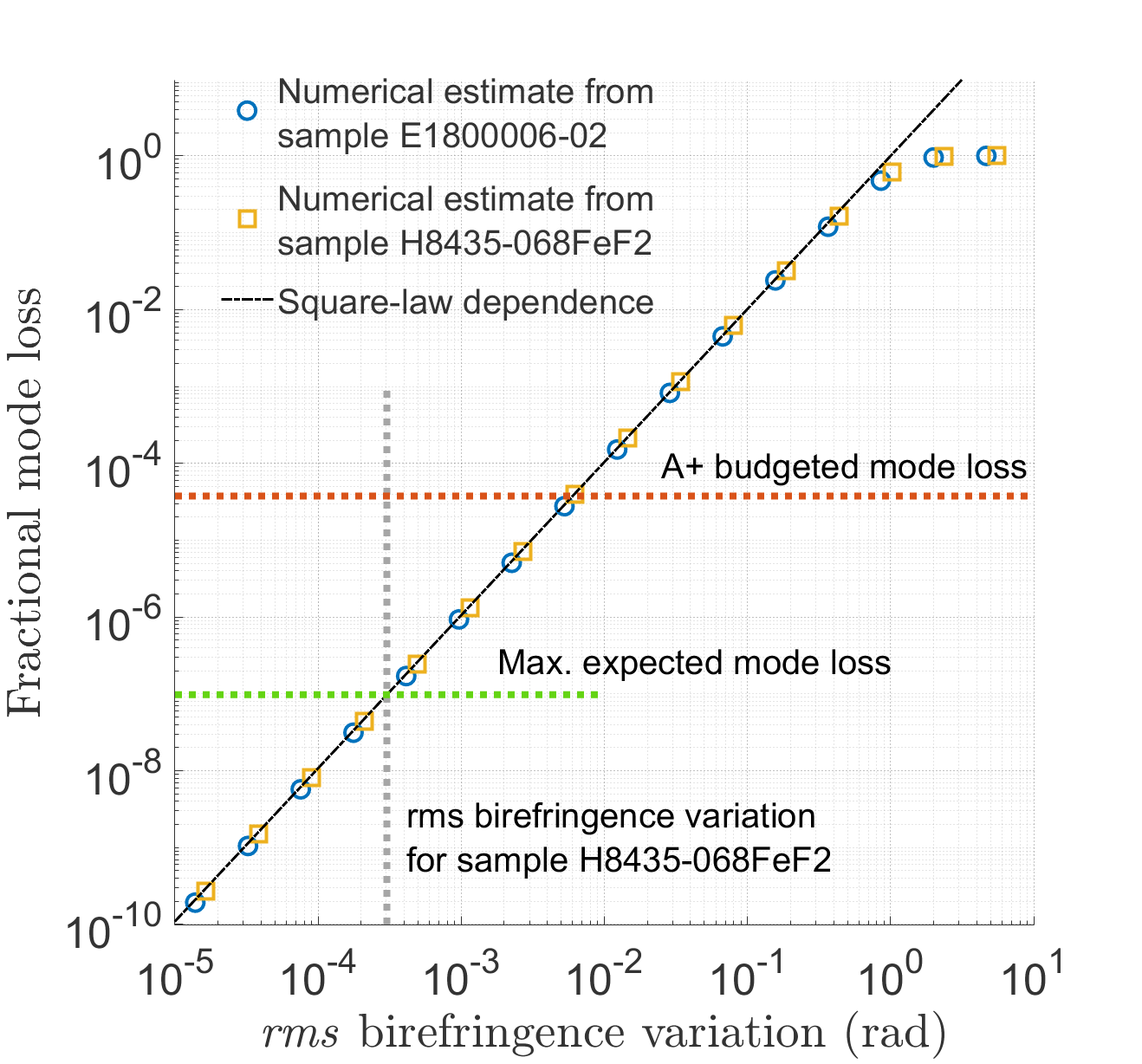}
    \caption{Power lost from the fundamental mode of the LIGO A+ arm cavity as a function of RMS birefringence variation. Birefringence maps for tiling onto a plane were taken from samples E1800006-02 and H8435-068FeF2. In the worst case scenario, using the map from H8435-068FeF2, the estimated mode loss is about two orders of magnitude below the total mode loss budgeted for A+. The rms level of birefringence non-uniformity for sthis sample is shown by the vertical dashed, grey line.}
    \label{fig:modeloss}
\end{figure}

The LIGO test-mass mirrors are significantly larger than the coatings measured here.  We mimicked a larger coating by tiling a plane with the data from a central 50~mm$\times$50~mm region of the 93~mm diameter coating (S/N: E1800006-02) shown in \fig{fig:10cm_variation_aligned}. The tiled plane was made sufficiently large (10 tiles by 10 tiles, 500~mm$\times$500~mm) that the amount of power in the beam tails falling outside the tiled region was negligible compared to the loss from scatter due to birefringence variation. The power lost to the fundamental mode was calculated by numerically evaluating the overlap integral between the $\mathrm{TEM_{00}}$ mode of the cavity and the perturbed mode with the phase variation impressed upon it by a single reflection. To check for numerical accuracy, we multiplied the birefringence map by a varying factor and found that the mode loss fraction had the expected square dependence on the amplitude of the birefringence (see e.g. Eq.~49 in reference \cite{Michimura2024}). The process was repeated for  similar 60~mm$\times$60~mm region near the lower left portion of the 194~mm coating (S/N: H8435-068FeF2) shown in \fig{fig:SampleA}. \fig{fig:modeloss} shows the result of the calculation. The mode loss is presented as a function of the RMS amplitude of the birefringence and the results from both birefringence maps are included on the same graph. For RMS birefringence under 1 mrad, which is satisfied by all the samples we measured, the power lost from the fundamental mode per reflection is at least one order of magnitude below the total optical loss per reflection  budgeted for the upcoming LIGO A+ detector~\cite{AplusDesignCurve}.

\section{Conclusions}

Our single-reflection method for measuring birefringence in large AlGaAs coatings has sufficient sensitivity to reveal spatial variations of less than 0.1~mrad.

The average birefringence measured by our method was consistent with the birefringence measured from resonance-peak splitting in optical cavities, together giving a birefringence $\psi = (1.09\pm0.18)$~mrad for a single reflection. The average birefringence is likely due to the Al$_{0.92}$Ga$_{0.08}$As layers which are under compression from the lattice mismatch.

Birefringence non-uniformity is primarily due to the bonding process. The bonding process leads to birefringence variation due to the inability of the coating to slide on the substrate as it bonds. It cannot relax from an initial curved state to the final flat state. To improve this, we need to ensure that the coating rolls onto the substrate in the manner of a rigid cylinder. Crystal defects also cause spatial variations in the birefringence but are limited to a small area around the defect. Non-uniform birefringence is associated with very small surface-height changes in the coating.

Optical scatter from a LIGO arm cavity due to the non-uniform birefringence seen in our measurements would cause single-reflection loss well under 1 ppm. Optical loss from non-uniform birefringence is therefore unlikely to contribute significantly to the total optical loss budget of a LIGO upgrade using AlGaAs coatings.

\section{Acknowledgements}
This work was performed under National Science Foundation grants NSF2110598 and NSF2409602. The authors would like to take the opportunity to thank Thorlabs Crystalline Solutions for their ongoing collaboration and support.

\bibliography{main}

\end{document}